\documentclass[aip,linenumbers,onecolumn,superscriptaddress]{revtex4}
\usepackage{graphicx}
\usepackage{epsfig}
\usepackage{bm}
\usepackage{amsmath}
\usepackage{amssymb}
\usepackage{wasysym}
\usepackage{epstopdf}
\usepackage{color}
\usepackage{xcolor}
\usepackage{epstopdf}
\usepackage{upgreek}
\usepackage{natbib}

\usepackage[colorlinks=false, bookmarks=true]{hyperref}

\newcommand{\etal}{\emph{et al.}}

\begin{document}


\title{How gravity and size affect the acceleration statistics of bubbles in turbulence} 
\author{Vivek N. Prakash}
\affiliation{ 
Physics of Fluids Group, Faculty of Science and Technology, J. M. Burgers Centre for Fluid Dynamics, University of Twente, P.O.Box 217, 7500 AE Enschede, The Netherlands.}

\author{Yoshiyuki Tagawa}
\affiliation{ 
Physics of Fluids Group, Faculty of Science and Technology, J. M. Burgers Centre for Fluid Dynamics, University of Twente, P.O.Box 217, 7500 AE Enschede, The Netherlands.}
\author{Enrico Calzavarini}
\affiliation{ 
Laboratoire de Mecanique de Lille CNRS/UMR 8107, Universite Lille 1 - Science et Technologie and PolytechÕLille, Cite Scientifique Av. P. Langevin 59650 Villeneuve dÕAscq, France.} 

\author{Juli\'an Mart\'inez Mercado }
\affiliation{ 
Physics of Fluids Group, Faculty of Science and Technology, J. M. Burgers Centre for Fluid Dynamics, University of Twente, P.O.Box 217, 7500 AE Enschede, The Netherlands.}

\author{Federico Toschi}
\affiliation{
Department of Physics, and Department of Mathematics and Computer Science, J. M.
Burgers Center for Fluid Dynamics, Eindhoven University of Technology, \\ 5600 MB Eindhoven, The Netherlands.\\
CNR-IAC, Via dei Taurini 19, 00185 Rome, Italy.}

\author{Detlef Lohse}
\affiliation{ 
Physics of Fluids Group, Faculty of Science and Technology, J. M. Burgers Centre for Fluid Dynamics, University of Twente, P.O.Box 217, 7500 AE Enschede, The Netherlands.}

\author{Chao Sun}
\affiliation{ 
Physics of Fluids Group, Faculty of Science and Technology, J. M. Burgers Centre for Fluid Dynamics, University of Twente, P.O.Box 217, 7500 AE Enschede, The Netherlands.}
\collaboration{\textit{International Collaboration for Turbulence Research}}

\date{\today}

\begin{abstract}
We report results from the first systematic Lagrangian experimental investigation in the previously unexplored regime of very light (air bubbles in water) and large particles ($D / \eta >> 1$) in turbulence. 
Using a traversing camera setup and particle tracking, we study the Lagrangian acceleration statistics of $\!\sim\! 3 \ mm$ diameter ($D$) bubbles in a water tunnel with nearly homogeneous and isotropic turbulence generated by an active-grid. The Reynolds number ($Re_{\lambda}$) is varied from 145 to 230, resulting in size ratios, $D / \eta$ in the range of 7.3--12.5, where $\eta$ is the Kolmogorov length scale. 

The experiments reveal that gravity increases the acceleration variance and reduces the intermittency of the PDF in the vertical direction. 
Once the gravity offset is subtracted, the variances of both the horizontal and vertical acceleration components are about $5 \pm 2$ times larger than the one measured in the same flow for fluid tracers. Moreover, for these light particles, the experimental acceleration PDF shows a substantial reduction in intermittency at growing size ratios, in contrast to neutrally buoyant or heavy particles. 
All these results are closely matched by numerical simulations of finite-size bubbles with the Fax\'en corrections. 
\end{abstract}

\maketitle

\section{\label{sec:level1}Introduction\protect\\ }

Suspensions of particulate materials, drops or bubbles carried by vigorously turbulent flows occur frequently both in the realm of natural phenomena (e.g. cloud formation) and in industrial applications (e.g. combustion in engines).
In order to quantify the statistical properties of such suspensions a prototype problem is often considered: the one of a dilute suspension of spherical particles in incompressible, statistically homogeneous and isotropic turbulence \cite{Toschi2009,LaPorta2001,Lohse2008}.
In this simple form the problem is defined by a set of three dimensionless parameters $\left[Re_{\lambda},  \Gamma, \Xi \right]$, respectively the Reynolds number based on the Taylor scale of the carrying flow,  the particle to fluid mass density ratio ($\Gamma \equiv \rho_{p} / \rho_{f}$) and the particle to dissipative-length ratio ($\Xi \equiv D/\eta$). Interesting theoretical questions concern how far the particle statistical properties  (e.g. moments, probability density functions (PDFs), correlations) of position, velocity and acceleration depart from the Lagrangian properties of the fluid.  The goal is to understand how such observables vary as a function of the control parameters.

Numerical studies have attempted to see how closely the approximate equations of Lagrangian dynamics -- which were known for a long time -- are able to capture the dynamics. 
Since full numerics (e.g. physalis \cite{Pro01}, front tracking \cite{Try01}) are too expensive for high $\mathrm{Re}_{\lambda}$, most simulations use a point particle model~\cite{Arneodo2008,Volk2008a,Biferale2008comp}, also known as the Maxey-Riley-Gatignol model \cite{max83,gat83}. In the real world, there are many situations where the particle size is larger than the Kolmogorov length scale of turbulence ($\Xi \gg 1$). Therefore, a considerable body of recent work (\cite{Voth2002,Qureshi2007,Qureshi2008, Calzavarini2009,Brown2009,Gibert2010,Volk2011}) has been dedicated to the characterization of these so-called \textit{finite-sized} particles. Numerical simulations with Fax\'en corrections to the point-particle approach~\cite{Calzavarini2009} correctly capture two important features from the experimental data for neutrally buoyant and heavy finite-sized particles \cite{Voth2002,Qureshi2007,Qureshi2008, Gibert2010,Brown2009,Volk2011}: the acceleration PDF of the finite-size particles in general show less intermittency than those of fluid tracers; and their acceleration variance decreases with increasing size ratio.
However, for light particles ($\Gamma<1$), the Fax\'en corrected numerics~\cite{Calzavarini2009} remarkably indicate an opposite trend for the acceleration statistics at growing $\Xi$: an initial increase of acceleration variance and intermittency, followed by a decrease. These predictions for light particles are awaiting experimental verification.  
However, these experiments are highly challenging in terms of the infrastructure needed, techniques and analysis, but are of great importance for a fundamental understanding of particles in turbulence. 

\begin{figure}
\centering
\includegraphics[width=0.45\textwidth]{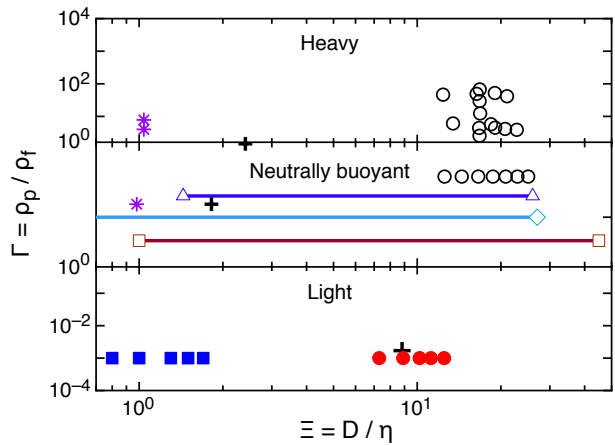}
\caption{Parameter space of the density ratio versus size ratio for particles in turbulence, from data available in literature: {\color{blue}$\triangle$} - Voth~\etal~\cite{Voth2002}, $\circ$ - Qureshi~\etal~\cite{Qureshi2008}, + Volk~\etal~\cite{Volk2008b}, {\color{cyan}$\diamond$} Brown~\etal~\cite{Brown2009}, {\color{purple}$\ast$} Gibert~\etal~\cite{Gibert2010}, {\color{brown}$\square$} Volk~\etal~\cite{Volk2011}, {\color{blue}$\blacksquare$} Martinez~\etal~\cite{Julian2012}, {\color{red}$\bullet$} Present work. 
Majority of previous studies have focused on $\Gamma\geq 1$, here we explore the $\Gamma \ll 1$ case.} 
\label{fig:phasediagram}
\end{figure}

In this paper we present the first systematic experimental investigation in the regime ($\Gamma\ll1, \Xi > 1$) of the parameter-space, i.e. very light and large particles (see Fig.~\ref{fig:phasediagram} for a summary of all the presently available measurements). For such an investigation we use air bubbles, which are dispersed in a turbulent water flow. We track these bubbles using a traversing camera system which can perform 2D recordings of the vertical and one horizontal component of the  bubble trajectories. The experiments are compared to numerical simulations based on the particle Lagrangian equations with  Fax\'en correction~\cite{Calzavarini2009}. We also experimentally study the effects of gravity, as it could be important at lower $\mathrm{Re}_{\lambda}$, and very few numerical studies~\cite{maz04} have taken gravity into account. 

The paper is organized as follows: in section~\ref{sec:exp}, we describe in detail the experimental setup (sec.~\ref{sec:setup}), the bubble deformability and size distribution (sec.~\ref{sec:def}). Next, we present results on the velocity statistics (sec.~\ref{sec:vel}) and the acceleration statistics (sec.~\ref{sec:acc}). The effects of gravity and size on the acceleration statistics are discussed in sections (sec.~\ref{sec:grav}) and (sec.~\ref{sec:size}), followed by the conclusions in section~\ref{sec:con}.

\begin{figure}
\centering
\includegraphics[width=0.5\textwidth]{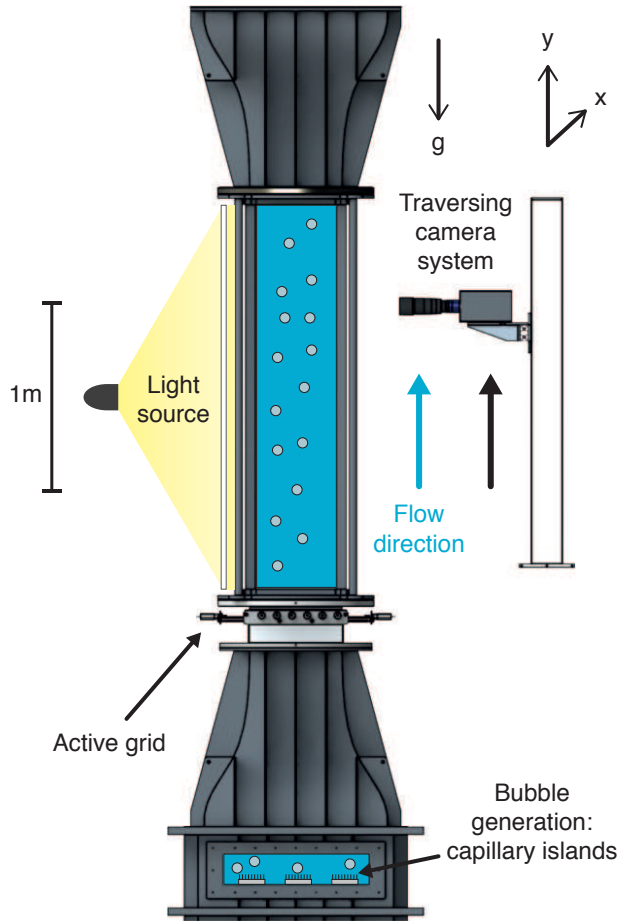}\\
\caption{The Twente Water Tunnel facility: vertical water tunnel with nearly homogeneous and isotropic turbulence generated by an active-grid. Bubbles are dispersed from below through capillary islands and the flow is in the upward direction. The camera moves upward at preset speeds, along with the bubbles, allowing the measurement of long-duration trajectories.}
\label{fig:twt}
\end{figure}

\section{Experiments and analysis}\label{sec:exp} 

\subsection{Experimental setup}  \label{sec:setup}

In our experiments, air bubbles in water, $\Gamma\approx10^{-3}$, are dispersed in a turbulent flow in the $8 m$ high Twente Water Tunnel (TWT) facility (see Fig.~\ref{fig:twt}(a)). Nearly homogeneous and isotropic turbulence is generated by the flow of water through an active grid~\cite{Poorte2002}. An optically transparent measurement section of dimensions 2$\times$0.45$\times$0.45 m$^3$ is located downstream of the active grid.  We recently reported results on $\sim 0.3\ mm$ sized bubbles (\cite{Julian2012,Yoshi2012}) using the same experimental facility (see also~ \cite{Martinez2010,Rensen2005}). Here, one important modification is made: the position of the active grid is switched from the position on top of the measurement section to its bottom. Furthermore, the direction of the water flow through the active grid is now upwards.  The bubbles are generated by blowing air through capillary islands (diameter 500 $\mathrm{\mu m}$) placed below the active grid. The bubbles rise through the measurement section along with the imposed water flow and escape through an open vent on top of the water tunnel. 
A surfactant (Triton X-100) is added to the tap water to reduce the bubble deformability~\cite{Takagi2011} (more details follow in section~\ref{sec:def}). 
 $\mathrm{Re}_{\lambda}$ is varied from 145--230 by changing the mean flow speed of water in the tunnel. The flow characterization is done using a hot-film probe placed in the center of the measurement volume~\cite{Julian2012}. The two-dimensional Lagrangian particle tracking experiments are carried out using a high-speed camera (Photron SA1.1) at an image acquisition rate of 5000 frames per second (fps) with a resolution of $768\times768$ pixels. The camera is focused on a 1-2 cm thick plane in the middle of the measurement section and the illumination is provided from the opposite side by a halogen light source placed behind a diffusive plate. The camera is mounted on a traverse system (Aerotech L-ATS62150 linear stage) which enables precise movement in the vertical direction at preset velocities. The detection of the bubble centers in the images is a non-trivial task because the bubbles often overlap or go out of focus (see Fig.~\ref{fig:twt}(b)). However, the circular Hough transform method~\cite{Taopeng} is successfully used to detect more than 90$\%$ of the bubbles (which are in focus) in the images. A 2D Particle Tracking code is then used to obtain the bubble trajectories over time. The data processing approach is the same as in Martinez~\etal~\cite{Julian2012}. The fitting window lengths ($N$) of the trajectory smoothing for the micro-bubble experiments (see Ref.~\cite{Julian2012} for more details) and present experiments have been chosen to be consistent. The fitting window length is $N=15$ for the present experiments at 5000 fps acquisition rate, and $N=30$ for the micro-bubble experiments at 10,000 fps acquisition rate.

\subsection{Bubble deformability and size distribution}  \label{sec:def}

A surfactant (Triton X-100) is added to the tap water to reduce the bubble deformability. The small amount of surfactant ($<$ 1 ppm) used in the present experiments is much below the critical micelle concentration, so the change in flow properties is negligible~\cite{Takagi2009}. 
Although the addition of surfactant reduces the surface tension by a few percent, 
it leads to another competing effect, namely the suppression of bubble coalescence at the source of injection. This second effect is more dominant and as a result we see a reduction of the bubble size~\cite{Takagi2011}.

The Fig.~\ref{fig:deform} below shows the difference between bubbles in tap water and in a surfactant solution. We clearly see a big difference in the shapes and sizes of the bubbles. The deformation is greatly reduced in the presence of surfactant, thanks to the smaller size. 
Nonetheless, in Fig.~\ref{fig:deform}, the bubbles still appear to be  deformed and anisotropic, but the effect is much less than without surfactants: 
The addition of surfactants leads to 
  a {\it reduced} deformability of the bubbles, not to a perfect spherical shape; the principal-axis deformation ratios are less than 1/2. The images in Fig.~\ref{fig:deform} are instantaneous snapshots from the experiments. It must be noted that the deformation of real bubbles is strongly time dependent, and the addition of surfactant significantly reduces this deformation over time. Thus, the addition of the surfactant renders the images amenable to processing, which is otherwise a nearly impossible task.

\begin{figure}
\centering
\includegraphics[width=0.45\textwidth]{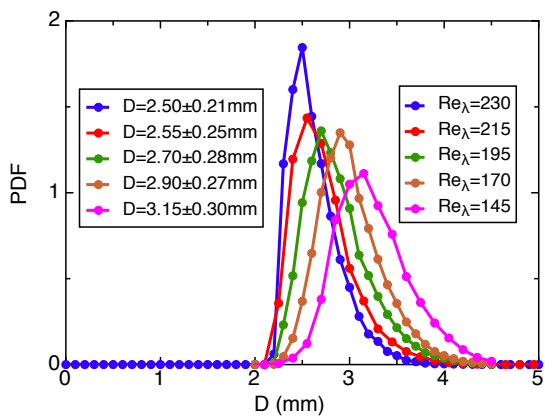}
\caption{Distribution function of bubble diameters, D, at different $\mathrm{Re}_{\lambda}$. }
\label{fig:sizedist}
\end{figure}

The bubble diameter has a weak dependence on $\mathrm{Re}_{\lambda}$, as shown in Fig.~\ref{fig:sizedist}. We select the peak values of the distribution as the characteristic bubble diameter; $D = 3.15, 2.90, 2.70, 2.55, 2.50$ mm (with absolute deviations $\sim\pm0.3$ mm, see Fig.~\ref{fig:sizedist}) at $\mathrm{Re}_{\lambda}$ = 145, 170, 195, 215, and 230, respectively. This variation in $\mathrm{Re}_{\lambda}$ changes the Kolmogorov length scale $(\eta)$, and the corresponding size ratios are $\Xi$ = 7.3, 8.9, 10.2, 11.2, and 12.5.

\begin{figure}
\centering
\includegraphics[width=0.7\textwidth]{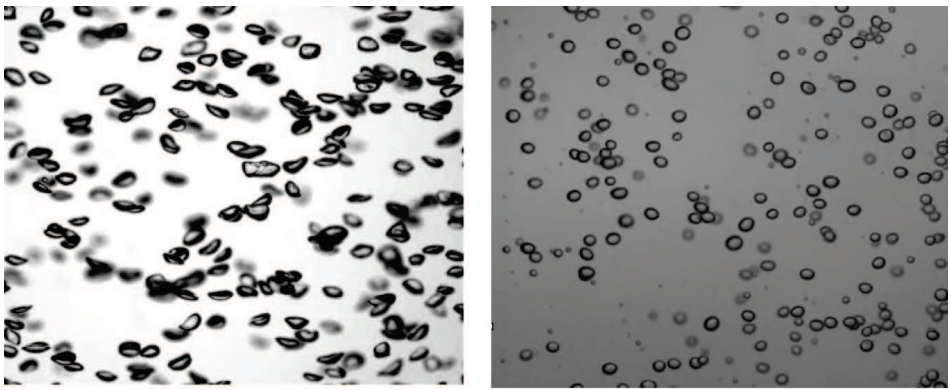}
\caption{Left panel: Rising bubbles of diameters $\sim 5 mm$ in tap water. Right panel: Bubbles of diameters $\sim 3 mm$ in surfactant solution (with $<$ 1 ppm Triton X-100). }
\label{fig:deform}
\end{figure}

%
\section{\label{sec:res}Results - Velocity Statistics} \label{sec:vel}

\begin{figure}[h!]
\centering
\includegraphics[width=0.4\textwidth]{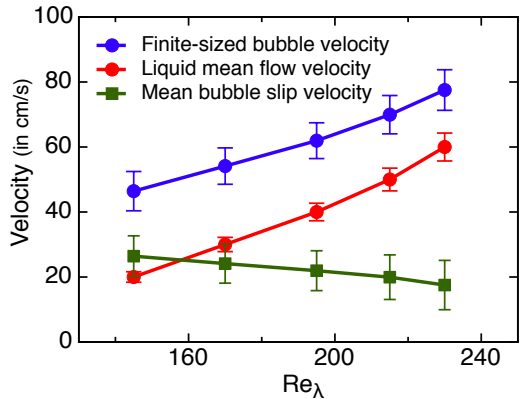}
\caption{Finite-sized bubble slip velocity at different $\mathrm{Re}_{\lambda}$. }
\label{fig:slip}
\end{figure}

In the present experiments, the bubble motion and liquid mean flow are both upwards. The terminal velocity of the $\sim3 mm$ bubbles in the absence of a liquid mean flow (still liquid) is determined from our experiments to be 25 $cm s^{-1}$. In the presence of a liquid mean flow, the bubbles attain their terminal velocity and then feel the turbulence of the surrounding liquid flow. In Fig.~\ref{fig:slip} we show the actual mean rise velocities of the bubbles (measured from the present experiments) compared to the liquid mean flow speeds. The bubbles experience a so-called `slip' and are able to rise with velocities higher than the liquid mean flow. We quantify this ``mean bubble slip velocity" by defining it as the difference between the bubble rise velocity and the liquid mean flow velocity. The variation of this mean slip velocity with $\mathrm{Re}_{\lambda}$ is shown in Fig.~\ref{fig:slip}. We see a decrease of the mean bubble slip velocity with increase in $\mathrm{Re}_{\lambda}$. This decreasing trend might be due to the decreasing bubble size at higher $\mathrm{Re}_{\lambda}$. 

\begin{figure}[h!]
  \centering
\includegraphics[width=0.9\textwidth]{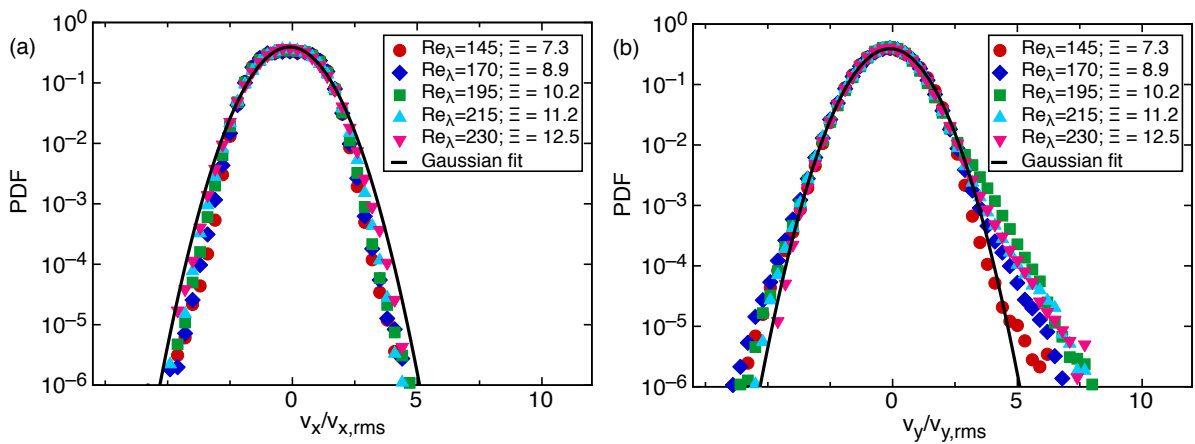}
  \caption{Velocity PDFs of the finite-sized bubbles at different $\mathrm{Re}_{\lambda}$ (and $\Xi$), (a) Horizontal component, $v_x$ (b) Vertical component, $v_y$ (c) flatness $F(v)$ versus $\mathrm{Re}_{\lambda}$. The solid line in (a) and (b) represents a Gaussian distribution fit. The velocity PDF in (a) is mostly sub-Gaussian and in (b) we see more intermittency than a Gaussian profile.}
  \label{fig:vPDF}
\end{figure}

We now present results on the probability distribution functions (PDFs) of the finite-sized bubble velocity obtained from the present experiments.
Fig.~\ref{fig:vPDF}(a, b) shows the PDF of the horizontal (x) and vertical (y) components of the normalized bubble velocity at the different $\mathrm{Re}_{\lambda}$ (and $\Xi$) covered in the present study. 
We observe that the x-component velocity distributions (Fig.~\ref{fig:vPDF}(a)) follow sub-Gaussian statistics (flatness values: $2.27-2.78$) and the y-component velocity distributions (Fig.~\ref{fig:vPDF}(b)) show a slightly higher intermittency compared to the Gaussian profile (flatness values: $2.89-3.77$). In previous experiments on finite-sized neutrally buoyant particles in turbulent von K\'arm\'an flows, Volk \etal~\cite{Volk2011} have obtained sub-Gaussian velocity distributions with flatness around 2.4$-$2.6, but in the past,  Gaussian-type flatness values (2.8$-$3.2) have also been reported~\cite{Voth2002}. The reason for deviations of the velocity distributions from Gaussianity is still an open question.



\section{\label{sec:res}Results - Acceleration statistics}  \label{sec:acc}
\subsection{Acceleration statistics: Gravity effect}  \label{sec:grav}
We first address the effect of gravity on the acceleration statistics. Since the buoyancy is proportional to bubble volume, while the laminar viscous drag grows with the linear size of the bubble, it is clear that for growing bubble sizes and fixed turbulence intensity (growing $\Xi$), buoyancy at some point shall dominate. The opposite is true for the case of fixed bubble size but increasing Reynolds numbers (again growing $\Xi$) in our experiments, where buoyancy loses its dominance at higher $\mathrm{Re}_{\lambda}$.   
In summary, it is acceptable to neglect the buoyancy force only for small bubbles $\Xi \lesssim 1$ \cite{Julian2012}, or for large $\mathrm{Re}_{\lambda}$, as we find later. 

One may expect that the buoyancy force will produce \textit{asymmetry}: on the vertical component statistics because buoyancy will add up to the upward acceleration events and will subtract from the downward events. However, we find that the asymmetry is almost negligible. First, we find that the mean value of vertical component of acceleration ($y$) is essentially zero, as it is for the horizontal one ($x$): $\langle a_x \rangle \simeq \langle a_y \rangle  = 0 \pm 0.2 g$, where $g$ is the acceleration due to gravity. 
Second, we observe that the probability density function (PDF) shape is only very weakly asymmetric; we indeed estimate the skewness, $S(a_i) \equiv \langle a_i ^3\rangle / \langle a_i^2\rangle^{3/2}$, and find $|S(a_x)| \leq 0.01$ and $|S(a_y)| \leq 0.1$. The skewness is comparable to the values found for $\Xi<2$ bubbles (from now on called microbubbles) studied in the same setup~\cite{Julian2012}. We can conclude that if any asymmetry is present it must be very weak.

\begin{figure}[h!]
\centering
\includegraphics[width=0.45\textwidth]{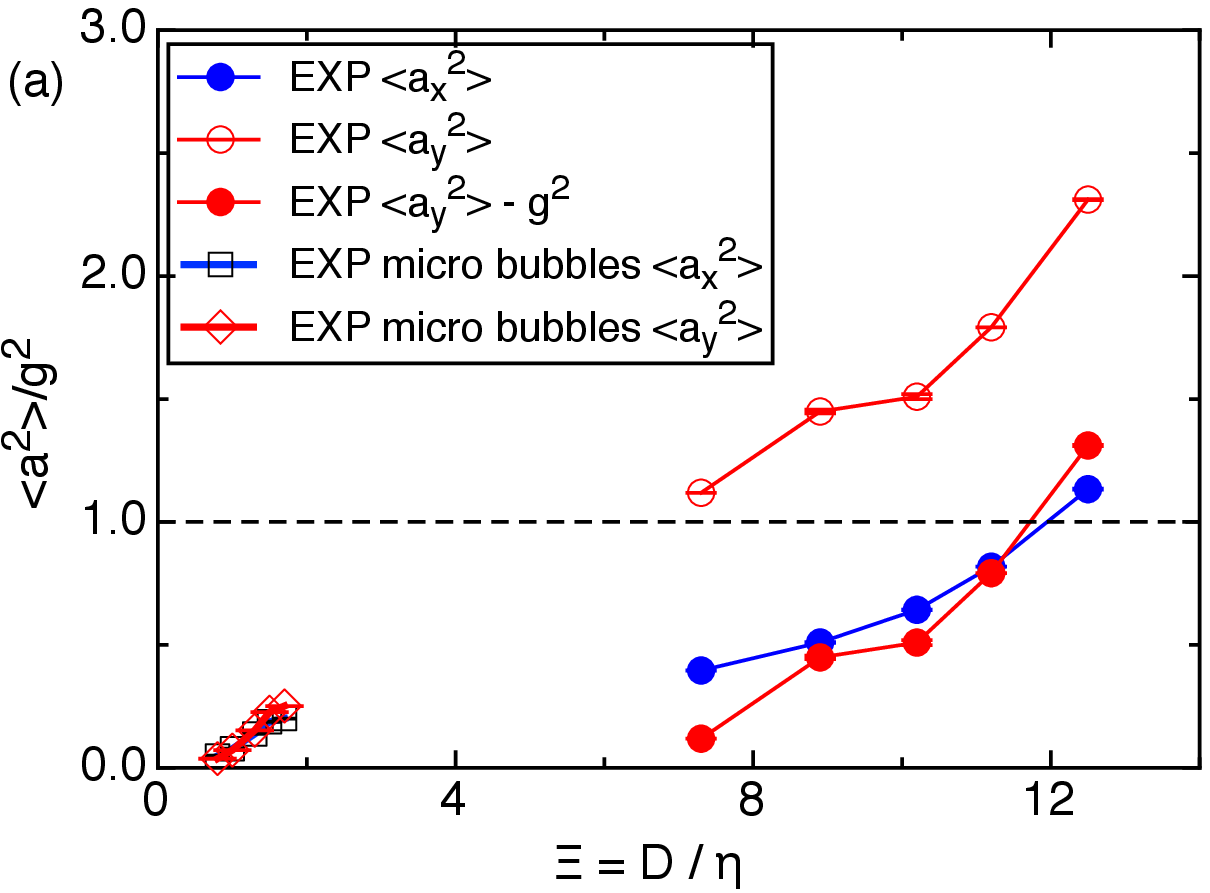}
\includegraphics[width=0.45\textwidth]{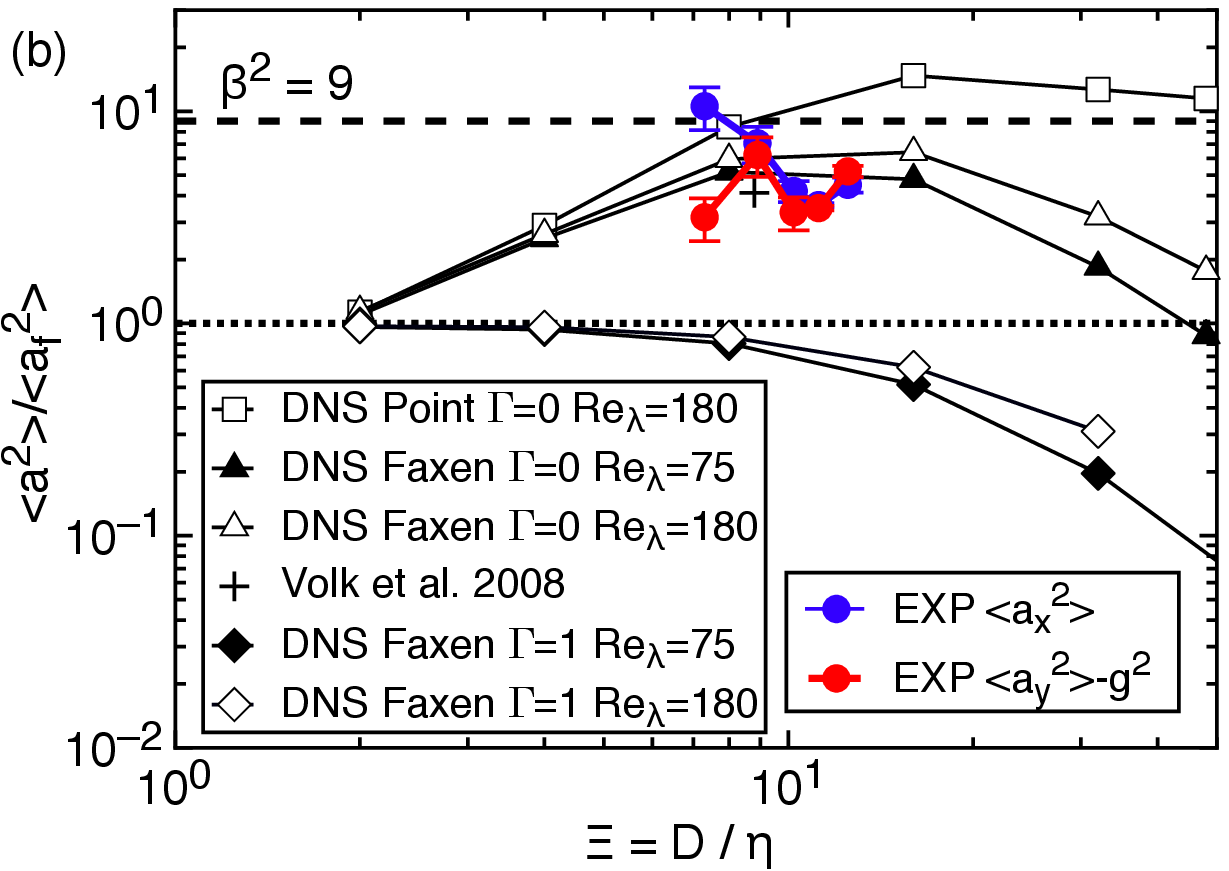}
\caption{(a) The single-component acceleration variance normalized by the gravitational acceleration for large bubbles and for micro-bubbles~\cite{Julian2012}. (b) The normalized acceleration variance versus the size ratio. \textbf{Legend :}  The present large bubble results ($\bullet$) are represented with errorbars, along with a gravity offset for the vertical component $a_y$. Open squares $\square$: point-particle DNS of bubbles ~\cite{Calzavarini2009} triangles $\vartriangle$ $\&$ $\blacktriangle$: bubbles from DNS with Fax\'en corrections~\cite{Calzavarini2009}, plus + : single experimental point for bubbles~\cite{Volk2008b}, diamonds $\lozenge$ $\&$ $\blacklozenge$: DNS of neutrally buoyant particles with Fax\'en corrections~\cite{Calzavarini2009}.} 
\label{fig:avar}
\end{figure}

The buoyancy force, however, produces a robust \textit{anisotropy}: different statistics for the vertical and the horizontal acceleration components. This influence of gravity is clearly visible in the second statistical moments of acceleration, shown in Fig~\ref{fig:avar}(a), where the  variance $\langle a_i^2\rangle$ is plotted. We find that $\langle a_y^2 \rangle \gg \langle a_x^2 \rangle$ for all $\Xi>7.3$ bubbles, while from the same figure it is evident that microbubbles have much closer variance values for the three cartesian components. 
If one calls $a_y'$ the vertical acceleration component in absence of gravity and assumes that $\langle a_y^2 \rangle = \langle (a_y' - g )^2 \rangle$, one gets $\langle a_y'^2 \rangle = \langle a_y ^2 \rangle - g^2$, since $\langle a_y' \rangle = 0$ because of isotropy. As can be seen from figure ~\ref{fig:avar}(a), this produces a reasonable collapse of the x-y data $\langle a_y^2 \rangle -g^2 \simeq \langle a_x^2 \rangle$. 
We emphasize that the collapse observed for the second order moment is non-trivial. It means that the statistical effect of gravity seems to be additive on the vertical direction with no effect on the horizontal component (see supplementary information for more discussions on this). An immediate consequence is that the effect of hydrodynamic forces coupling different cartesian directions, as for example the lift force, turns out to be unimportant here. 

 \begin{figure}[htp!]
 \centering
  \includegraphics[width=0.8\textwidth ]{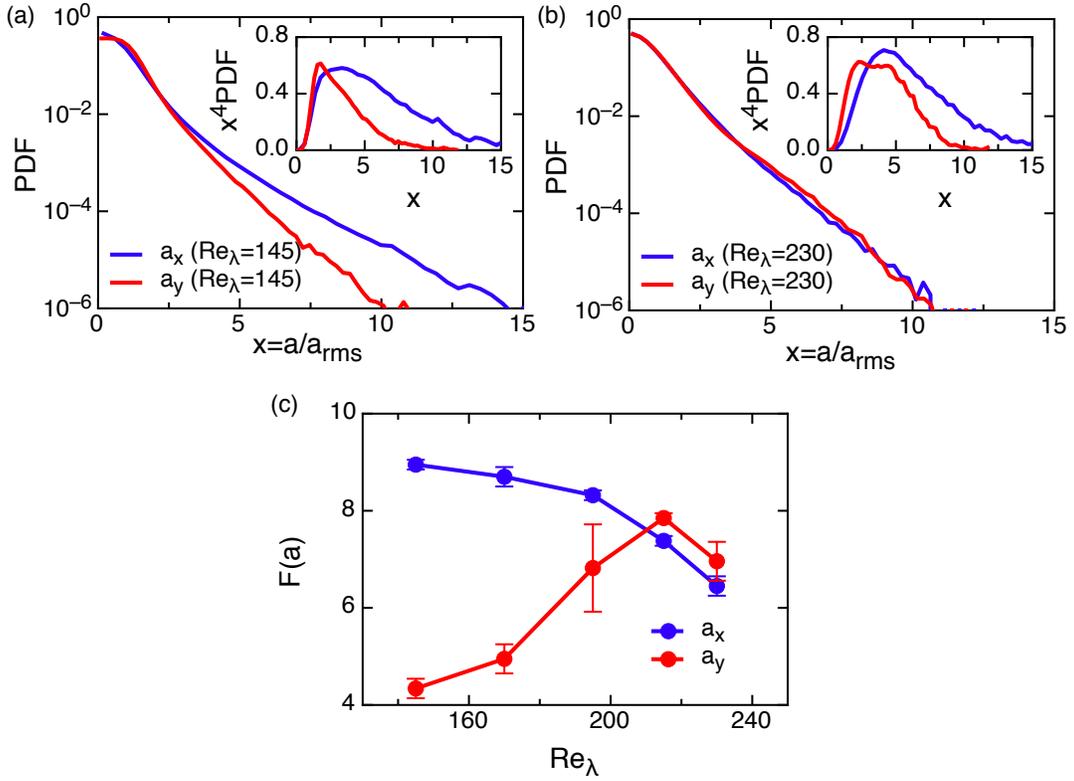}
  \caption{Normalized bubble acceleration PDFs of $a_x$ and $a_y$ components at  $\mathrm{Re}_{\lambda} = (a) 145$ and (b) $230$, which are respectively the smallest and the largest $\mathrm{Re}_{\lambda}$ considered in our experiments. As expected, at the lower $\mathrm{Re}_{\lambda}$ the tails of the normalized PDF (in other words the flatness) of the vertical component are reduced compared to the horizontal one, while at the higher $\mathrm{Re}_{\lambda}$ the anisotropy is negligible. The insets in (a) and (b) show the fourth-order convergence test. (c) The acceleration flatness for the present bubbles (with errorbars) versus $\mathrm{Re}_{\lambda}$. }
  \label{fig:aPDF}
\end{figure}

As mentioned above, we have found that the acceleration variance $\left<a_y^2\right>$ in the vertical direction is augmented by an offset (or a correction factor) that depends on $g^2$. Although we have found that the offset of $\sim g^2$ seems to work well for the present finite-sized bubbles, we do not yet have a physical understanding of the origin of this correction. When one examines the equations governing the motion of bubbles in a turbulent flow (Maxey-Riley-Gatignol equations)~\cite{max83,gat83} including gravity; the corresponding acceleration variance offset would then be $\sim 4g^2$, which does not work for the present data. 
Clearly, there seems to be a complex interplay between gravity and inertia, and more experimental and numerical work is needed before we can arrive at solid conclusion on the gravity correction factor. These corrections will be explored in detail in future work.


The finite-sized bubble acceleration PDFs and the flatness values at two selected cases of $\mathrm{Re}_{\lambda}$ are shown in Fig.~\ref{fig:aPDF}. The quality of statistics in the present experimental data is seen in the insets of Fig.~\ref{fig:aPDF}(a,b), which demonstrate that the fourth-order moments are well converged. Hence, we can calculate the flatness directly from the distribution without resorting to a fitting procedure (e.g. as in Volk~\etal~\cite{Volk2011}). In Fig.~\ref{fig:aPDF}(c), we observe that the flatness values of $a_y$ at $\mathrm{Re}_{\lambda}$ $\le195$ are less than those of $a_x$. At higher $\mathrm{Re}_{\lambda}$ ($\ge 215$), the flatness values of $a_x$ and $a_y$ become comparable, and the corresponding PDFs show a nice collapse in Fig.~\ref{fig:aPDF}(b). The decreased intermittency in the vertical component $a_{y}$ for the low $\mathrm{Re}_{\lambda}$ (seen in Fig.~\ref{fig:aPDF}(a)) is due to gravity. Apparently, there are two regimes: at lower $\mathrm{Re}_{\lambda}$, gravity has an effect on the acceleration statistics in the vertical direction, which is no longer the case at higher $\mathrm{Re}_{\lambda}$.

\subsection{Acceleration Statistics: Size effect} \label{sec:size}

Once the statistical influence of buoyancy has been disentangled, we now study how the gravity-less acceleration variance changes at increasing $\Xi$, in other words we study the purely hydrodynamic size-effect on particles which are lighter than the surrounding fluid.
It is convenient to look at the relative change of the bubble acceleration variance with respect to fluid tracers $ \langle a_{i,f}^2  \rangle$. For $a_{i,f}$  we use here the acceleration of the microbubbles as found in Ref.~\cite{Julian2012}, because given their small size, $\Xi<2$, they behave almost like Lagrangian tracers. Furthermore, the $\mathrm{Re}_{\lambda}$ numbers studied in the present experiments are very close to the ones analyzed in Ref.~\cite{Julian2012} and the flow conditions are the same. Fig.~\ref{fig:avar}b reports such a normalized acceleration variance $\langle a_i^2\rangle/\langle a_{i,f}^2 \rangle$ versus the size ratio $\Xi$.
We see that both acceleration components reach a level of $5\pm2$ times the variance measured in the same flow for fluid tracers. They are also in agreement with the previous single experimental datapoint of Ref.~\cite{Volk2008b}. 
To have an interpretation of this measurement we compare it with the results from numerical simulations at similar $\mathrm{Re}_{\lambda}$.
Fig.~\ref{fig:avar}b reports results from two different types of Lagangian particle simulations, first the so called point-particle (PP) simulation that only takes into account the hydrodynamic effects of added mass and Stokes-drag, and a second simulation which adds on Fax\'en corrections (FC) to the mentioned terms~\cite{Calzavarini2009,Calzavarini2012}. 
The PP model predicts for $\Xi \to \infty$ an asymptotic limit of normalized acceleration variance which is 9 times that for tracers (as a result of the dominance of the added-mass term). 
From Fig~\ref{fig:avar}b, we deduce that for $\Xi \approx 10$ in the PP model, it is about 15 times the value for tracers, reflecting that we are not yet in the asymptotic limit. For larger $\Xi$ the results from the PP model indeed seem to approach the asymptotic value of normalized acceleration variance of 9. 
Fax\'en corrections to the added mass term reduce the value at $\Xi = 10$ which is about 7. Fig~\ref{fig:avar}b clearly shows that numerical simulations of Fax\'en corrected finite-sized bubbles show a good agreement with the experimental measurements (contrary to the PP model which overestimates the result). The present results are the first systematic measurements to confirm that the acceleration variance increases with finite-size for bubbles, matching the FC numerical simulations~\cite{Calzavarini2009}. Note again that the trend is very different for the case of neutrally buoyant particles and heavy particles, for which the normalized acceleration variance is always $\le1$ ~\cite{Volk2011,Calzavarini2009} (Fig.~\ref{fig:avar}b).



\begin{figure}
\centering
\includegraphics[width=0.45\textwidth]{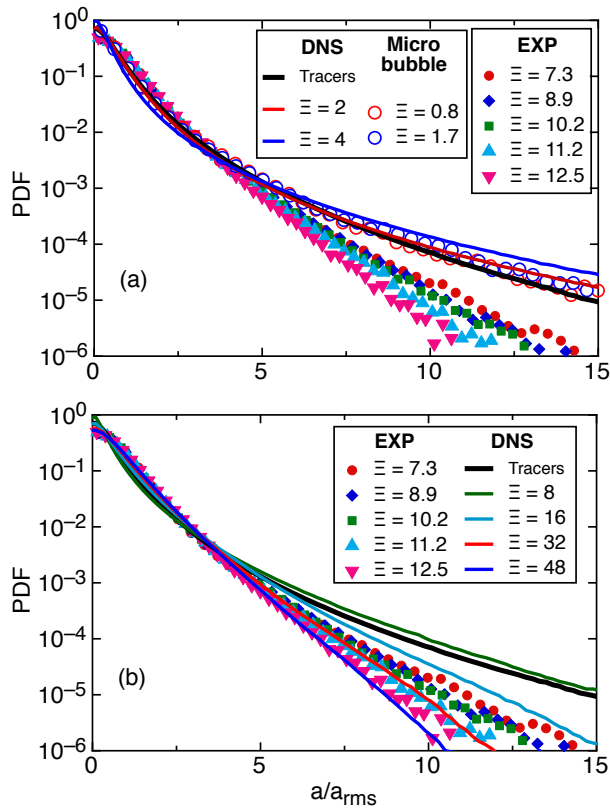}
\caption{The normalized acceleration PDFs of the horizontal component $|a_x|$, from the present bubble experiments (EXP), microbubble measurements~\cite{Julian2012}, and numerical simulations (DNS) with the Fax\'en model at $\mathrm{Re}_{\lambda}$ = 180 \cite{Calzavarini2009}. (a) Comparison with results for point-like particles; DNS Tracers, microbubbles, and Fax\'en DNS of bubbles for $\Xi$ $\le$ 4. (b) Comparison with Fax\'en DNS of larger bubbles, $\Xi$ $\ge$ 8.}
\label{fig:Flatness}
\end{figure}

Making statements on the fourth order statistical moments of acceleration based on experimental measurements is a delicate endeavor, see e.g. the detailed analysis by Volk \textit{et al.}~\cite{Volk2008b}. 
To study the finite-size effects on the bubble acceleration, in the following we will focus only on the PDF shape of the normalized $x-$acceleration, $a_x / \langle a_x \rangle_{\textrm{rms}}$, which is not directly affected by gravity.
In Fig.~\ref{fig:Flatness}(a) first we plot such a curve for microbubbles~\cite{Julian2012} and see that its shape falls on the one for fluid tracers, and on $\Xi<2$ bubbles, from DNS at similar $\mathrm{Re}_{\lambda}$. This is an evidence of both the above mentioned passive nature of microbubbles and of the similarity between turbulence realized in the Twente Channel flow and the one produced in homogeneous and isotropic DNS.
In the same panel we see that, in sharp contrast, the finite-sized bubbles show a strongly reduced intermittency. It is the first time that such a substantial change in intermittency at growing size ratios is experimentally observed: neither for solid neutrally buoyant particles \cite{Brown2009,Volk2011}, nor for heavier bubbles \cite{Qureshi2008} was it detected before.

In order to clarify further the magnitude of such an effect, in Fig.~\ref{fig:Flatness}(b) we compare the acceleration PDF from the DNS simulations with Fax\'en corrections and the present experiments. 
There is an initial increase of flatness in the DNS simulations for $\Xi \lesssim 8$, which probably reflects the limitations of the Fax\'en corrections. 
For $\Xi\gtrsim7-8$ both the numerics and the experiments show a significant reduction of the tails of the PDF. However, the DNS appears to underestimate its functional behavior by approximately a factor 2-3 in the size parameter $\Xi$.
The reason for this discrepancy is presently unclear. 
One reason could be that the simulations neglect the two way coupling and are just approximated in the implementation of Fax\'en terms.
Another possible reason for the discrepancy which deserves further study is the deformability of real bubbles. The deformation process absorbs/releases energy from/to the turbulent environment, a process which may have an effect on acceleration statistics. These issues motivate further investigations to better understand light particles in turbulence.

\section{Conclusion} \label{sec:con}
We performed measurements of Lagrangian acceleration in the previously unexplored regime of large (compared to $\eta$) and very light (with respect to $\rho_f$) particles in turbulence. Bubbles of size $\!\sim\! 3 \ mm$ diameter ($D$) were tracked in turbulent flow conditions in a water tunnel. The explored range of Reynolds number ($Re_{\lambda}$) and size ratios, $\Xi = D / \eta$ were 145--230, and 7.3--12.5, respectively.

Gravity produces anisotropy in the acceleration statistics of the vertical component -- it adds a $g^2$ offset  to the variance, and decreases the intermittency of PDF, at lower $\mathrm{Re}_{\lambda}$. It was found that the interaction between gravity and inertia is complex, and this deserves further study.

The acceleration variances and the intermittency clearly indicate the finite-size effect, and the results are in good agreement with DNS simulations with the Fax\'en corrections. To improve the current understanding, in the future we plan to study rigid hollow spheres in the turbulent water flow, and to vary their diameters, $D$, at fixed $\mathrm{Re}_{\lambda}$ numbers. 

\begin{acknowledgments}
We thank G. Voth, L.-P. Wang, H. Xu and B. Luethi for discussions, G.-W. Bruggert, M. Bos and B. Benschop for help with the experimental setup. We also thank J. van Nugteren and B. Colijn for assistance with the experiments. J.M.M. acknowledges support from the Foundation for Fundamental Research on Matter (FOM) through the FOM-IPP Industrial Partnership Program: \emph{Fundamentals of heterogeneous bubbly flows}. We also acknowledge support from the European Cooperation in Science and Technology (COST) Action MP0806: \emph{Particles in turbulence}. Finally, we thank two anonymous referees for their constructive suggestions.
\end{acknowledgments}

\end{document}